\documentclass[journal]{IEEEtran}

\usepackage{graphics}
\usepackage{graphicx}
\usepackage{caption}
\usepackage[numbers,sort&compress]{natbib}
\usepackage{amsfonts}
\usepackage{amsmath}
\usepackage{url}
\usepackage[caption=false]{subfig}
\usepackage[table,xcdraw]{xcolor}
\usepackage{multirow}
\usepackage{enumitem}
\usepackage[hidelinks]{hyperref}
\usepackage{cleveref}
\crefformat{footnote}{#2\footnotemark[#1]#3}
\usepackage{tikz,xcolor}

\definecolor{lime}{HTML}{A6CE39}
\DeclareRobustCommand{\orcidicon}{%
	\begin{tikzpicture}
	\draw[lime, fill=lime] (0,0) 
	circle [radius=0.16] 
	node[white] {{\fontfamily{qag}\selectfont \tiny ID}};
	\draw[white, fill=white] (-0.0625,0.095) 
	circle [radius=0.007];
	\end{tikzpicture}
	\hspace{-2mm}
}

\foreach \x in {A, ..., Z}{%
	\expandafter\xdef\csname orcid\x\endcsname{\noexpand\href{https://orcid.org/\csname orcidauthor\x\endcsname}{\noexpand\orcidicon}}
}

\begin{document}

\title{A Secure Fog Based Architecture for Industrial Internet of Things and Industry 4.0}

\author{Jayasree~Sengupta\orcidA{},
        Sushmita~Ruj,~\IEEEmembership{Senior~Member,~IEEE,}
        and~Sipra~Das~Bit,~\IEEEmembership{Senior~Member,~IEEE}
\thanks{J. Sengupta and S. Das Bit are with the Department
of Computer Science \& Technology, Indian Institute of Engineering Science \& Technology, Howrah, India, E-Mail: (jayasree202@gmail.com ; sdasbit@yahoo.co.in).}
\thanks{S. Ruj is with CSIRO, Data61, Australia and Cryptology and Security Research Unit, Indian Statistical Institute, Kolkata, India, E-Mail: (sushmita.ruj@gmail.com ; sushmita.ruj@csiro.au).}
\thanks{Manuscript submitted : 15 July, 2019 ; Revised : 26 December, 2019 and 28 March 2020 ; Accepted : 12 May 2020.}
}

\markboth{IEEE Transactions on Industrial Informatics}%
{Sengupta \MakeLowercase{\textit{et al.}}: A Secure Fog Based Architecture for Industrial Internet of Things and Industry 4.0}

\maketitle

\begin{abstract}
    The advent of Industrial IoT (IIoT) along with Cloud computing has brought a huge paradigm shift in manufacturing industries resulting in yet another industrial revolution, Industry 4.0. Huge amounts of delay-sensitive data of diverse nature are being generated which needs to be locally processed and secured due to its sensitivity. But, the low-end IoT devices are unable to handle huge computational overheads. Also, the semi-trusted nature of Cloud introduces several security concerns. To address these issues, this work proposes a secure Fog-based IIoT architecture by suitably plugging a number of security features into it and by offloading some of the tasks judiciously to fog nodes. These features secure the system alongside reducing the trust and burden on the cloud and resource-constrained devices respectively. We validate our proposed architecture through both theoretical overhead analysis and practical experimentation including simulation study and testbed implementation.
\end{abstract}


\section{Introduction}

Industrial IoT (IIoT) refers to the use of certain IoT technologies including various smart objects capable of monitoring, collecting, analyzing and making intelligent choices in an industrial setting without human intervention for the betterment of its environment \cite{i2}. With the emergence of the fourth industrial revolution, Industry 4.0 specially emphasizes on the manufacturing industry scenarios. Industry 4.0 is focused on digitizing and integrating all physical processes across the entire organization \cite{i5}. Thus, it is a major paradigm shift from the previous centrally controlled automated machines and processes into a more decentralized manufacturing. This enables the smart factories to effectively handle the growing complexities and situations like complete shutdown of assembly lines while simultaneously carrying out production more efficiently \cite{i3}.

Both IIoT and Industry 4.0 are attempts of making the industrial systems more scalable, robust and faster in terms of response as well as production. However, for effectively realizing these systems to its full potential, extending the features of the cloud closer to the end devices for locally processing tasks in a timely manner is an essential requirement. Smart industrial equipments are resource constrained and performing costly computations would drain them out quickly. Moreover, computations in the cloud has its own drawbacks such as large response time, disruptions in the underlying communication networks and data security and privacy issues \cite{i1}. One of the solutions to these problems is to introduce a middleware such as fog which can perform most of the tasks locally instead of getting them processed either in the end devices or remote server/cloud. Fog Computing maybe defined as \textit{``an extension of the cloud computing paradigm that provides computation, storage, and networking services between end devices and traditional cloud servers"} \cite{i4}. Primarily, fog computing brings in the advantages of scalability, better Quality of Service (QoS), agility, efficiency and decentralization. Apart from these the potential of fog computing is widespread as it reduces latency thereby reducing cost, preserves network bandwidth and also provides better security \cite{i6}.

\noindent We observe from the above that introducing fog as a middleware in IIoT to eliminate the drawbacks brought about by the cloud-centric architecture will potentially benefit the industries. The major contributions of our work are:

\begin{itemize}[leftmargin=*]
    \item We propose our secure Fog-based IIoT architecture by suitably plugging a number of important security features into it.
    \item The proposed architecture reduces the overhead on low-end devices and the latency in decision making alongside eliminating trust on the cloud.
    \item We validate the architecture with a theoretical overhead analysis and practical experimentation including simulation and testbed implementation.
\end{itemize}

\noindent The remainder of the paper is structured as follows. Section II highlights the Related Work. Section III describes our proposed Secure Fog-based IIoT architecture. Section IV establishes the proposed design. Section V highlights the performance evaluation of our architecture. Finally, Section VI concludes our work.

\section{Related Work}

This section provides a brief overview on : (a) the proposed IIoT based architectures and (b) the IIoT based security schemes proposed in literature.

The works \cite{r1,r2} have focused on proposing improved architectures for IIoT, specifically with respect to Industry 4.0. Wan et al. \cite{r1}, have proposed a Software-defined IIoT architecture to introduce greater flexibility in the network. However, Software Defined Networking (SDN) based IIoT architecture has a few drawbacks like, standardizing SDN is still under process, therefore standardization of SDN based IIoT is a complex task. SDN-based IIoT acts as a centralized controlling system which introduces latency in data forwarding. To reduce this problem of latency, the works \cite{a1,r2} have proposed the use of fog as a middleware in IIoT between the industrial devices and the cloud to promote distributed controlling of network traffic. However, their proposed architecture haven't considered any security measures to protect the data exchanged within the system and is therefore susceptible to various kinds of security threats and attacks like tampering.

The works \cite{r4,r5,r8} have proposed privacy preserving, lightweight or blockchain-based authentication schemes to achieve mutual authentication between various users and/or devices within the IIoT systems. In the context of data authentication, \cite{r6,r7} have proposed Certificateless Signature (CLS) schemes for IIoT systems to achieve the said objectives. The work \cite{r9} has proposed an Ethereum blockchain-enabled mechanism using Deep Reinforcement Learning (DRL) to create a reliable IIoT environment. Furthermore, Huang et al. \cite{r10} have proposed a credit-based proof-of-work (PoW) mechanism for IIoT devices to protect sensitive data confidentiality.

From the above discussions it is clear that preserving the security and privacy of the system as well as the data in transit have been less investigated. Further, decentralizing the entire network traffic to bring computation to the edge while ensuring security is another concern. At the same time reducing latency in real time data processing is also essential. This motivates us to design a secure Fog-based IIoT architecture addressing the aforementioned issues. 

\section{Secure Fog-Based IIoT Architecture}
In this section, we propose our Secure Fog-based IIoT Architecture in the context of Industry 4.0. But before that we discuss the major needs/requirements for incorporating security features in this architecture. 

\subsection{Security and Privacy Requirements}
The major requirements are as follows:

\begin{itemize}[leftmargin=*]
    \item The devices at the perception layer (e.g. industrial robots) and the fog nodes  belong to different vendors. Thus, it is extremely important to build a trust relationship between them by authenticating each other.
    \item Unauthorized access to either the data from the perception layer devices or the computing resources of the fog nodes should be prevented.
    \item Preserving integrity of raw data in transit and thereby preventing tampering is another important concern. 
    \item The cloud is considered to be semi-trusted or at times malicious, therefore it is extremely important to preserve confidentiality of data stored in the cloud. Also retaining the control over usage of such data is a concern. 
    \item Reducing the computational burden on the resource-constrained perception layer devices is a necessity. 
    \item Finally, securely performing computations on encrypted data stored in the cloud is another challenge . 
\end{itemize}

To include each of the aforementioned security requirements in an industrial setting, we primarily propose our Secure Fog-based IIoT architecture.

\subsection{Our Proposed System Architecture}
\begin{figure} [!ht]
\begin{center}  
\includegraphics[width=3.2in, height=2.5in]{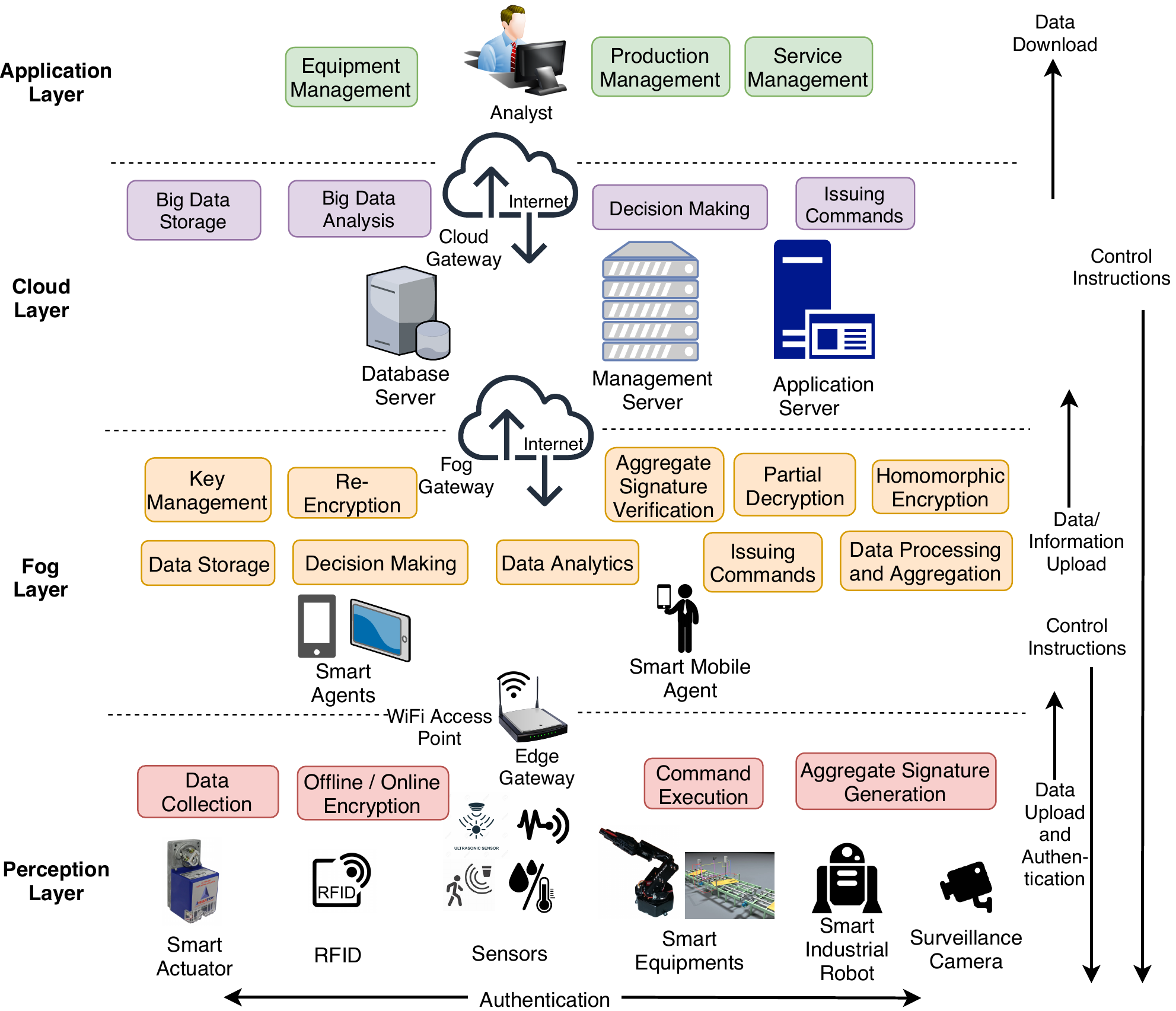}
\caption{\small \sl Secure Fog-based IIoT Architecture \label{fig:Image1}}
\end{center}  
\end{figure}

The architecture is shown in Figure \ref{fig:Image1} and is composed of four layers, namely the perception layer, fog layer, cloud layer and application layer.

\noindent \textbf{Perception Layer:} This layer consists of industrial devices equipped with different kind of sensors (e.g. temperature), RFID tags and surveillance cameras as per requirement. It also consists of smart equipments (e.g. robots) required to automate industrial processes, transport raw goods etc. The main functions of this layer include data collection, command execution and authentication of devices as well as data (using signatures). The raw data collected by IoT devices in this layer are transferred to the Fog layer using some edge gateway like Wifi Access points. 

\noindent \textbf{Fog Layer:} The fog layer is the additional layer we have introduced in our proposed architecture which comprises of fog nodes. Fog nodes are devices that necessarily reside close to the edge of the network and are often the first point of contact to the IoT end-devices \cite{a1}. To implement this, the fog layer may comprise of smart devices like smartphones, tablets, laptops, etc. or even virtual servers and smart mobile agents like humans carrying smartphones/tablets depending on the need of the application as mentioned in Section II.

\noindent We have introduced the fog layer in our architecture, to primarily achieve three main objectives : (a) Minimizing the latency in decision making by extending the cloud closer to where the devices are (b) Offloading the major computations from the resource constrained end devices and/or honest but curious cloud servers to the more trustworthy fog nodes and (c) Preserving the battery life of the resource-constrained IoT devices for a longer period by offloading the intensive computations. In our proposed architecture, the processing of the defined security features are distributed between the different layers in the following manner: the tasks which are to be executed by the IoT devices only for successful implementation of the defined security features are performed in the perception layer. Rest of the tasks which are flexible enough to be performed either by the IoT devices/proxy servers in Cloud or by the fog nodes are offloaded to the fog layer. This mechanism relieves the trust and burden on the cloud and resource-constrained devices respectively.
    
\noindent The IoT devices that sense raw data forward it to their nearest fog node. This fog node then performs either of the following two actions depending on the type of data it has received. If the data is highly time-sensitive, the fog node analyzes it and sends a control command back to the respective device. On the other hand, if the data is less time-sensitive, it is forwarded to the cloud for long term storage and/or big data or historical analysis. The thousands of fog nodes along the industrial floor also send periodic summaries of the data collected by them to the cloud for further analysis purposes. In this way the fog layer effectively bridges the gap between the IoT devices and the cloud and also performs some major computationally intensive tasks for securing the system on behalf of both of them. According to our setup, out of all the fog nodes, a few of them may behave as a proxy as per requirement.

\noindent \textbf{Cloud Layer:} The cloud layer consists of different kinds of servers (e.g. database server) performing various kind of activities. The cloud is responsible for storing huge chunks of data and performing other tasks (e.g. big data analysis). Based on these, the cloud issues and forwards instructions to the fog nodes. In our proposed secure Fog-based IIoT architecture, data is generally stored in the cloud in encrypted format. 

\noindent \textbf{Application Layer} For an industrial setting, the application layer consists of users (e.g. production manager) working on smart terminals to manage the entire workflow of the industry. The users download required data from the cloud using Internet which acts as the cloud gateway. Based on the analysis performed on these data, an intelligent decision is made to improve the overall quality of production, service or equipment management in the industry.

\section{Proposed Design}

A wide range of security vulnerabilities exists in the context of IIoT and Industry 4.0. Therefore, in this design, we choose an existing security scheme for each of these features and propose suitable modifications to it to enhance the security and performance of our proposed architecture. A couple of exemplary industrial scenarios where these features are essential to apply are discussed in \textit{Appendix A}. A common bilinear map is used for most of the below-mentioned schemes. Two groups $\mathbb{G}_1$ and $\mathbb{G}_T$ of large prime order \textit{q} is chosen where $g$ is the generator of $\mathbb{G}_1$. The pairing is defined as a map $\hat{e} : \mathbb{G}_1 \times \mathbb{G}_1 \rightarrow \mathbb{G}_T$. The detailed description about bilinear maps and pairing concepts are available in \textit{Appendix B}.

\subsection{Access Control and Authentication for Devices}

Authentication and access control are two of the most important concerns in an industrial IoT setting. Authentication between devices within the same layer as well as devices across layers should be incorporated in an industrial setting to build a trust relationship between such devices. Moreover, only trusted fog nodes should be allowed to access IoT device's data, or even issue commands to an IoT device which brings access control into light. 

The IoT devices are generally resource constrained and authentication/access control requires a lot of cryptographic operations. Therefore, the major computationally intensive operations need to be outsourced to the fog nodes, in order to save battery life of IoT devices. The work \cite{c4} has proposed an Attribute Credential Based Public Key Cryptography (AC-PKC) scheme for a fog based IIoT architecture which can be used both for authentication as well as access control. Therefore, we adopt their scheme in our proposed architecture to achieve its benefits and completely secure our architecture. 

\subsection{Secure Data Aggregation}

\begin{figure} [!ht]
\begin{center}  
\includegraphics[width=3in, height=1.6in]{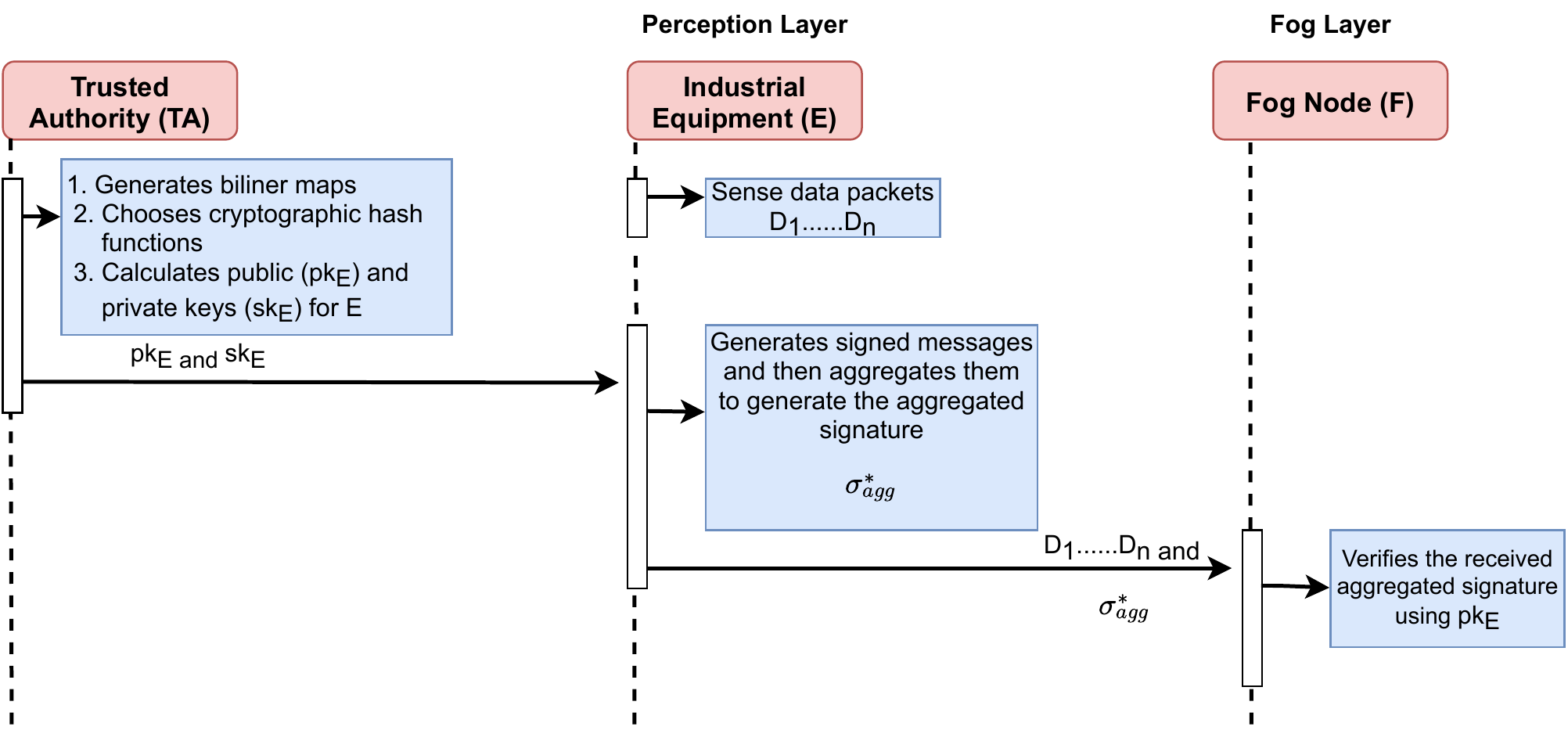} 
\caption{\small \sl Sequence Diagram for Secure Data Aggregation \label{fig:Image2_1}}
\end{center}  
\end{figure}

The sensory data that are transferred to the fog nodes for analysis can get tampered midway leading to drastic effects like severe damage to a machinery or even complete shutdown of an assembly line. To deal with such unprecedented conditions it is extremely important to protect the confidentiality and integrity of the data (raw data/commands). Signing the data (i.e. digital signature) not only ensures that the data is authenticated by the sender but also signifies that the data is not tampered and is a good choice in such cases. BLS Signature scheme \cite{c2} is one of the widely used signature schemes. However, if every frame is to be signed, the size of the signature becomes long. Aggregate BLS signature \cite{c2} is shorter in size and has a property that size of the aggregate signature is constant irrespective of the number of signatures being aggregated.  Let us assume that an industrial equipment (E) in the perception layer is sending multiple data packets ($D_1,\ ...,\ D_n$) within a short frame to a fog device (F). Figure \ref{fig:Image2_1} demonstrates in detail the exact steps performed by each of the entities in our proposed architecture to implement Aggregate BLS and the interactions between them.

\noindent \textbf{Setup:} The setup phase is performed by a Trusted Authority (TA) who generates bilinear map as defined in Section IV. TA also chooses a cryptographic hash functions $H_1:\{0,1\}^* \rightarrow \mathbb{G}_1$. All the public parameters are published and the private key ($sk=x \in \mathbb{Z}_q$) and public key ($pk=g^x$) for each individual device is calculated. Thus, E acquires its public key ($pk_E$) and private key ($sk_E$) respectively.

\noindent \textbf{Sign:$\ Sign(M_j,sk_i)\rightarrow \sigma$.} The sign phase is performed by E when it wishes to send data packets to F. For a group of packets $D_1,\ ...,\ D_n$, each of the individual data packets and private key ($sk_E$) is taken as input. The sign function, then returns the signed messages $\sigma_1\ ...\ \sigma_n$ which are calculated as $\sigma_j=H_1{(D_j)}^{sk_E}$

\noindent \textbf{Aggregate: $\ Aggregate(\sigma_{agg}(D_1,\ ...,\ D_n),\sigma,M)\rightarrow \sigma^*_{agg}$}. Aggregation is also performed by E as 
$\sigma^*_{agg}=\prod_{j=1}^n\sigma_j$.

\noindent \textbf{Verify:} F receives  a group of data packets $D_1\ ...\ D_n$ along with the aggregated signature (${\sigma^*}_{agg}$). F then verifies the signature, using the public key of E ($pk_E$) as follows:
$$\prod_{j=1}^n \ \hat{e}(pk_E,H_1(D_j))=\hat{e}(\sigma^*_{agg},g)$$

\noindent If the verification takes place successfully, the fog node is assured that all the received data is original (i.e. not tampered). In this way, the objective is achieved with minimum computational overhead as well as reduced latency which is extremely suitable for an IIoT environment.

\subsection{Secure Data Sharing with Multiple Users}

In an industrial environment, huge amounts of sensitive data needs to be outsourced to the cloud for sharing among various users for analysis purposes. The traditional approaches of securely sharing data only with authorized users (e.g. asymmetric encryption) have several limitations (e.g. huge computation and bandwidth cost to the data owner) \cite{c1}. Moreover, sharing the same data with multiple users require the sender to encrypt the data individually for each user. This would become a tedious and computationally intensive task for the sender. To achieve end to end data confidentiality during data sharing with authorized users, re-encryption of data is a promising choice. Re-Encryption is a mechanism which allows a proxy (typically a third party) to alter (i.e. re-encrypt) an encrypted message from a sender in such a way so that it can easily be decrypted by the recipient using his/her own private key \cite{c1}. Moreover, the message remains secret from the proxy. Even if the encrypted data needs to be shared with multiple users, the sender just needs to generate multiple re-encryption keys. Using these key, the proxy would re-encrypt the message for each individual user thereby drastically reducing the computational burden on the sender. 

\begin{figure} [!ht]
\begin{center}  
\includegraphics[scale=0.36]{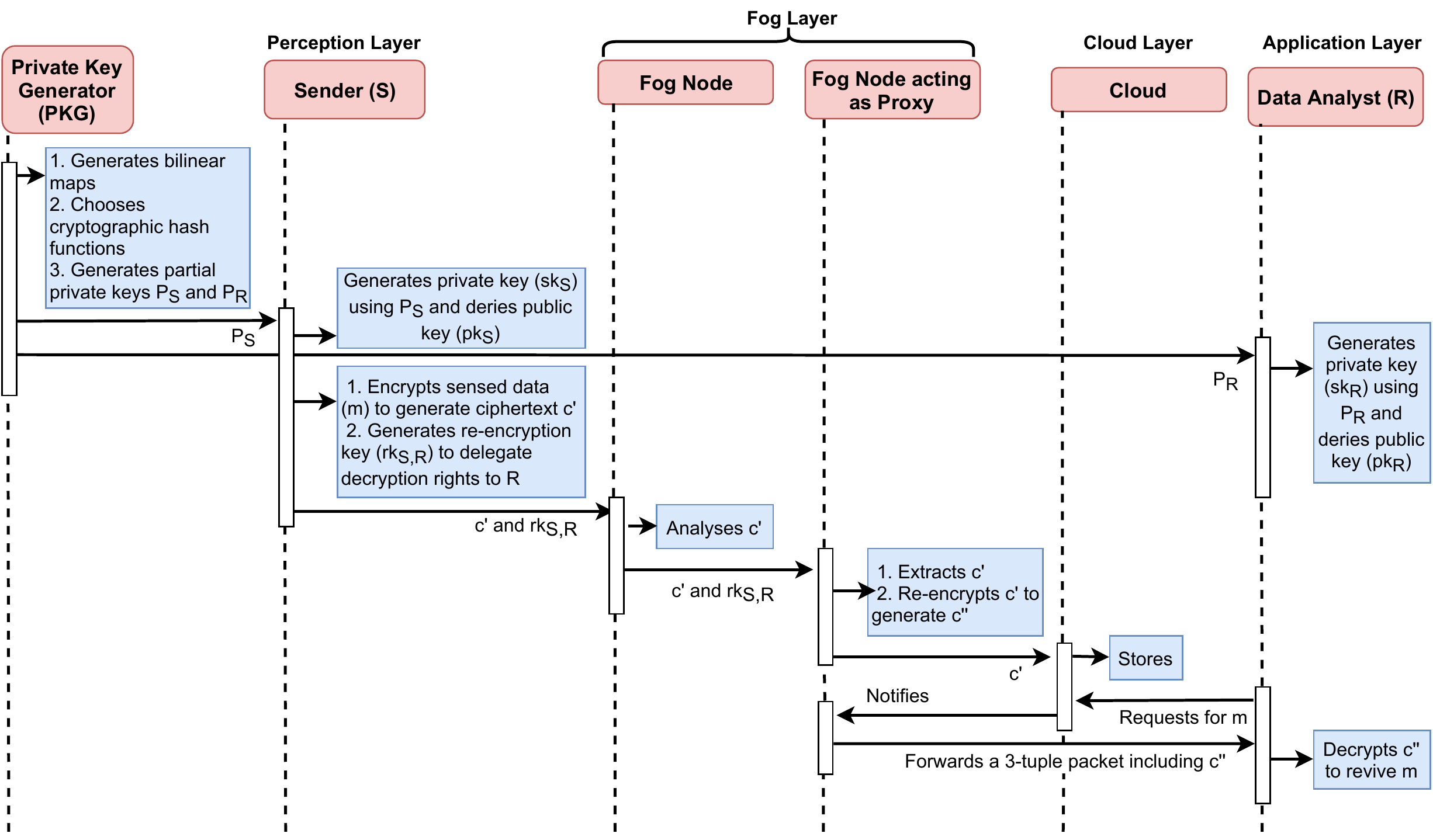}
\caption{\small \sl Sequence Diagram for Secure Data Sharing with Multiple Users \label{fig:Image3_1}}
\end{center}  
\end{figure}

The existing re-encryption schemes in literature \cite{c1,e1,e2,e3} utilize the cloud for key management and encryption-based access control apart from data storage. We modify the Certificateless Proxy Re-Encryption (CL-PRE) \cite{c1} scheme to eliminate trust on the Cloud. We achieve this by delegating the key management and re-encryption task to the fog nodes to fit into our proposed secure Fog-based IIoT architecture. Figure \ref{fig:Image3_1} demonstrates in detail the exact steps performed by each of the entities within our proposed architecture and the interactions between them. Let us assume that an industrial equipment in the perception layer acts as the sender (S) and there is a data analyst (R) at the industry, acting as a user. \smallskip

\noindent \textbf{PKG Setup:} The Private Key Generator (PKG) generates a bilinear map as defined in Section IV. PKG also chooses two cryptographic hash functions $H_1:\{0,1\}^* \rightarrow \mathbb{G}_1$ and $H_2:\mathbb{G}_T \rightarrow \mathbb{G}_1$. Finally, it randomly picks an integer, $mk \in \mathbb{Z}^{*}_q$ as master key to publish $g^{mk}$. For each of its members the PKG issues and forwards a partial private key as $P_i = {g_i}^{mk}$ ($g_i$ is calculated as per \cite{c1}). S and R receive their own partial private keys as $P_S$ and $P_R$ respectively.

\noindent \textbf{Key Generation:} Each member then randomly chooses a secret integer $k_i \in \mathbb{Z}^{*}_q$. Using $P_i$ and $k_i$, the members then generate their private keys as per \cite{c1}. Thus, both S and R derive their private keys as $sk_S$ and $sk_R$ respectively. Similarly, S and R also derive their public keys as $pk_S$ and $pk_R$ respectively (according to \cite{c1}). For decryption delegation, S chooses a random integer $d$, the tuples $<sk_S,d>$ are kept secret whereas $g^d$ is published as a part of $pk_S$.

\noindent \textbf{Encryption:} Whenever S has some sensed data (say $m$), and it needs to provide decryption delegation to R, S randomly chooses $r$. The encryption to generate a ciphertext ($c'$) takes place as follows:\smallskip
$$ c'= C'_S(m) = (g^{dr}, g^r, m.\hat{e}({g}_S^r,g^{d.k_S})) = (c_0,c_1,c_2)$$

\noindent \textbf{Re-Encryption Key Generation:} For delegating decryption capabilities to the receiver (R), the sender (S) randomly chooses $y \in \mathbb{G}_T$ and computes the re-encryption key ($rk_{S,R}$) as follows:
$$ rk_{S,R} = (g_S^{-sk_S}.H_2^d(y),C_R(y))=(c_4,C_R(y))$$

\noindent where $C_R(y) = g^r, m.\hat{e}({g}_R^r,g^{sk_R})$. S then forwards a 2-tuple $<c', rk_{S,R}>$ packet to the nearest Fog Node. The fog node on receiving this 2-tuple packet, analyzes it and decides whether the data needs to be forwarded to the cloud or not. If the data is meant for the cloud, then  it forwards the packet to another fog node acting as a proxy. Otherwise, it acts on the data itself and provides a feedback to S. 

\noindent \textbf{Re-Encryption by Fog:} The Fog node acting as a proxy, on receiving $<c',\  rk_{S,R}>$ packet, performs two tasks. Firstly, it extracts $c'$ and stores it in the cloud. Next, it re-encrypts $c'$ under $rk_{S,R}$ to generate $c''$ as follows:

$$ c'' = c_2. \hat{e}(c_4,c_1) = m.\hat{e}(H_2^d(y),c_1)$$

\noindent Whenever R requests for this specific data, the cloud notifies the respective Fog node from which it had received the encrypted data. This particular Fog node then forwards a 3-tuple $<c_0, c'', C_R(y)>$ packet to R. 

\noindent \textbf{Decryption:} R decrypts $C_R(y)$ as explained in \cite{c1} to retrieve $y$. The plaintext message can then be revived by computing 
$ c''/\hat{e}(H_2(y),c_0) = m $

\noindent This method reduces the trust dependency on the cloud by delegating the key management and re-encryption task to the fog nodes. Furthermore, fog nodes being a property of the industry, are more trustworthy compared to the cloud.

\subsection{Fine-Grained Access Control with Improved Performance}

As Cloud Service Providers (CSPs) can be malicious, data in encrypted form needs to be stored in the cloud. However, encryption and decryption requires complex operations which is often quite expensive for resource constrained devices. To handle such situations and simultaneously incorporate fine-grained access control, fast offline/online encryption and outsourced partial decryption is introduced. Partial Decryption is a phenomenon in which the data requester generates a transformed version of the decryption key. This transformed key allows a proxy (typically an untrusted third party) to partially decrypt the ciphertext in a way so that it cannot gain any information about the original message. This partially decrypted ciphertext can then be fully decrypted by the data requester without performing any costly operations. 

\begin{figure} [!ht]
\begin{center}  
\includegraphics[width=\columnwidth, height=2.2in]{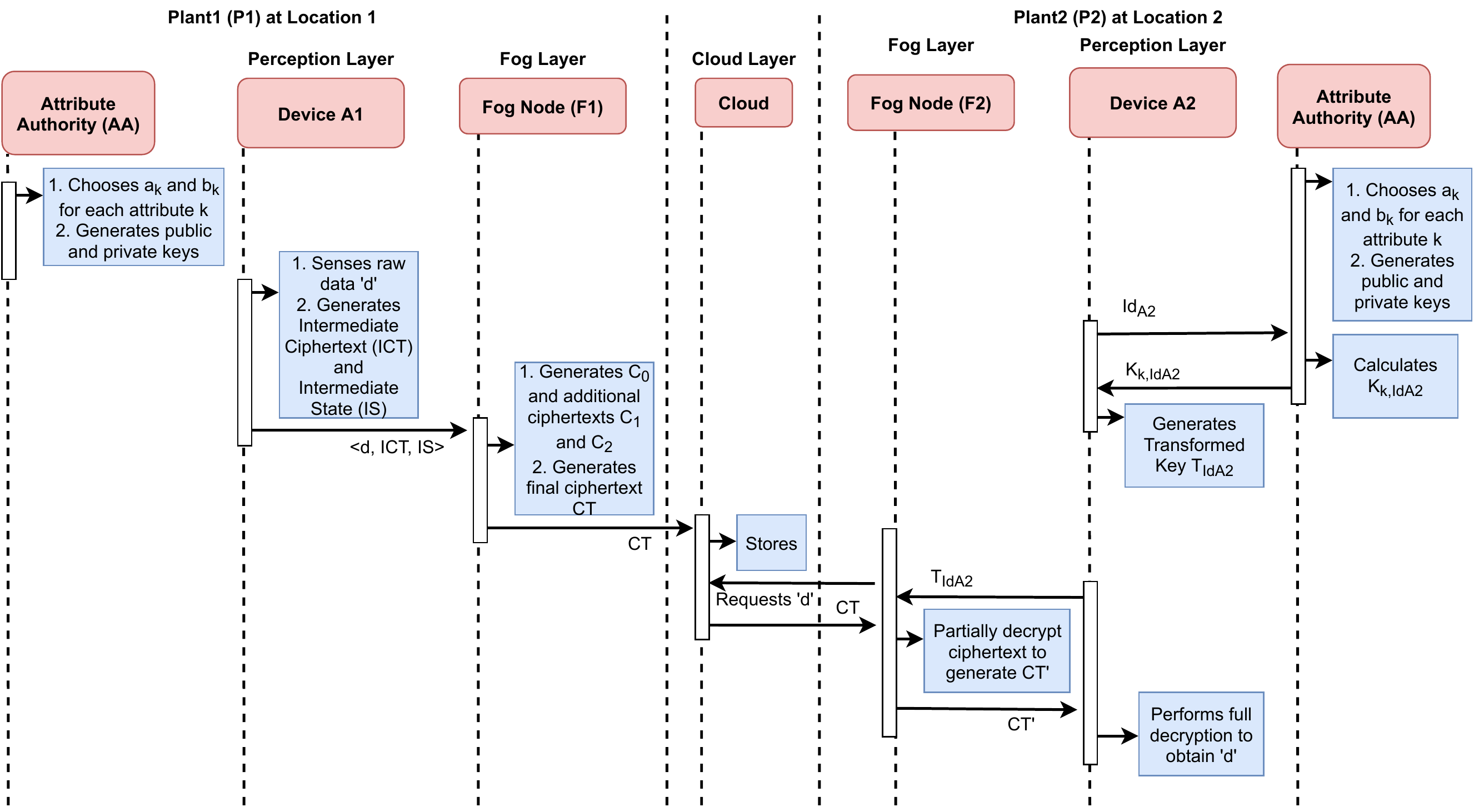}
\caption{\small \sl Sequence Diagram for Fine-Grained Access Control with Improved Performance \label{fig:Image4_1}}
\end{center}  
\end{figure}

The existing partial decryption schemes in literature \cite{c3,e4,e5,e6} utilize the cloud for outsourced decryption. We modify one of the existing schemes \cite{c3}, to promote the aforementioned advantages as well as completely eliminate the need of trusting the cloud by introducing the fog layer. Figure \ref{fig:Image4_1} illustrates the exact steps to implement fine-grained access control using partial decryption in our proposed architecture. Let us assume that an industry has two plants ($P_1$ and $P_2$ respectively) located at two distant places. A perception layer device (say $A_1$) in $P_1$ is the data owner, who uploads data to the cloud and another perception layer device (say $A_2$) in $P_2$ is the data requester querying the same data. \smallskip

\noindent \textbf{Global Setup:} This phase generates a bilinear map as defined under Section IV. Every device in the perception layer has a unique identifier ($Id$) and a cryptographic hash $H_1$ which maps $Id$ to elements in $\mathbb{G}_1$.

\noindent \textbf{Authority Setup:} There are more than one Attribute Authority (AA) in the system who controls different user attributes. For each attribute $k$ belonging to AA, it randomly chooses $a_k, b_k \in \mathbb{Z}_q$ to generate $pk$ and $sk$ as per \cite{c3}. Therefore, an AA ($i$) in $P_2$ controlling device $A_2$ publishes $pk_i$ as its public key and keeps $sk_i$ secret. \smallskip

\noindent The following two phases are performed at $P_1$, which acts as the data generator (as per our assumption).

\noindent \textbf{Intermediate Encryption:} Whenever a raw/sensory data (say $d$) is generated by $A_1$ it performs this phase. $A_1$ chooses $x=1$ to $X$ (where $X$ is the total number of attributes attached to $d$) and uses the global parameters to generate the intermediate ciphertext (${ICT}_{x \in [1,X]}$) and its intermediate state (${IS}_{x \in [1,X]}$). A 3-tuple packet $<d,ICT,IS>$ is then forwarded to the nearest fog node ($F_1$) [ICT and IS is calculated as per \cite{c3}]. 

\noindent \textbf{Full Encrypt:} The fog node ($F_1$) on receiving this 3-tuple packet uses the global parameters and the relevant public keys to generate the final ciphertext ($CT$). It randomly chooses $m_s \in \mathbb{Z}_q$ to generate $C_0=d\hat{e}(g,g)^{m_s}$. The additional cipher texts $C_1$ and $C_2$ are calculated according to \cite{c3}. The final ciphertext is then calculated as below and stored in the cloud.
$$CT=((\mathbb{A},\rho^*), C_0,{\{ICT,C_1,C_2,IS\}})_{x \in [1,X]}$$
where $\mathbb{A}$ is a $l\times m$ access matrix with $\rho$ mapping its rows to attributes \cite{c3}.\smallskip

\noindent The following phases are executed at $P_2$ where device $A_2$ has requested to access the data ($d$) generated by $A_1$.

\noindent \textbf{Key Generation:} This phase is run by AA within whose range $A_2$ falls. For an attribute $k$, this node calculates and sends $K_{k,Id_{A2}}=g^{a_k}H_1({Id_{A2}})^{b_k}$ to $A_2$ (having $Id_{A2}$).

\noindent \textbf{Key Transformation:} This phase is executed by $A_2$, where it chooses a random number $r \in \mathbb{Z}_q$ to generate the transformed key ($T_{Id_{A2}}=(K^{1/r}_{k,Id_{A2}},H_1({Id_{A2}})^{1/r}$). This transformed key is then forwarded to the nearest fog node ($F_2$) which downloads the requested data from the cloud.

\noindent \textbf{Partial Decryption:} $F_2$ on receiving the transformed key ($T_{Id_{A2}}$) and the downloaded ciphertext ($CT$), uses the global parameters to partially decrypt the ciphertext to produce $CT'=\{CT_1,CT_2\}$ (as per \cite{c3}), which is then sent to $A_2$. 

\noindent \textbf{Full Decryption:} $A_2$ computes $CT'_2=CT^{1/r}_2$ and $CT=CT'_2.CT_1$. Then, it computes $CT^r=\hat{e}{(g,g)}^{m_s}$ to obtain $d=C_0/CT^r$.

\noindent The computationally intensive tasks (e.g. partial decryption) are performed by the fog node ($F_2$). This completely eliminates trust on the cloud and also reduces the computational burden on the low-end perception layer devices.

\subsection{Secure Computation}

To protect sensitive data from honest but curious cloud servers, data is encrypted. However, this setting brings in a new challenge of performing computations on encrypted data. Traditionally, fully homomorphic encryption was used as solution which is not a practical choice. In order to reduce the trust on cloud servers and burden on the users, secure computation is delegated to the fog nodes in our architecture. These nodes then perform computations on the encrypted data stored in the cloud.

\begin{figure} [!ht]
\begin{center}  
\includegraphics[width=3.2in, height=2.4in]{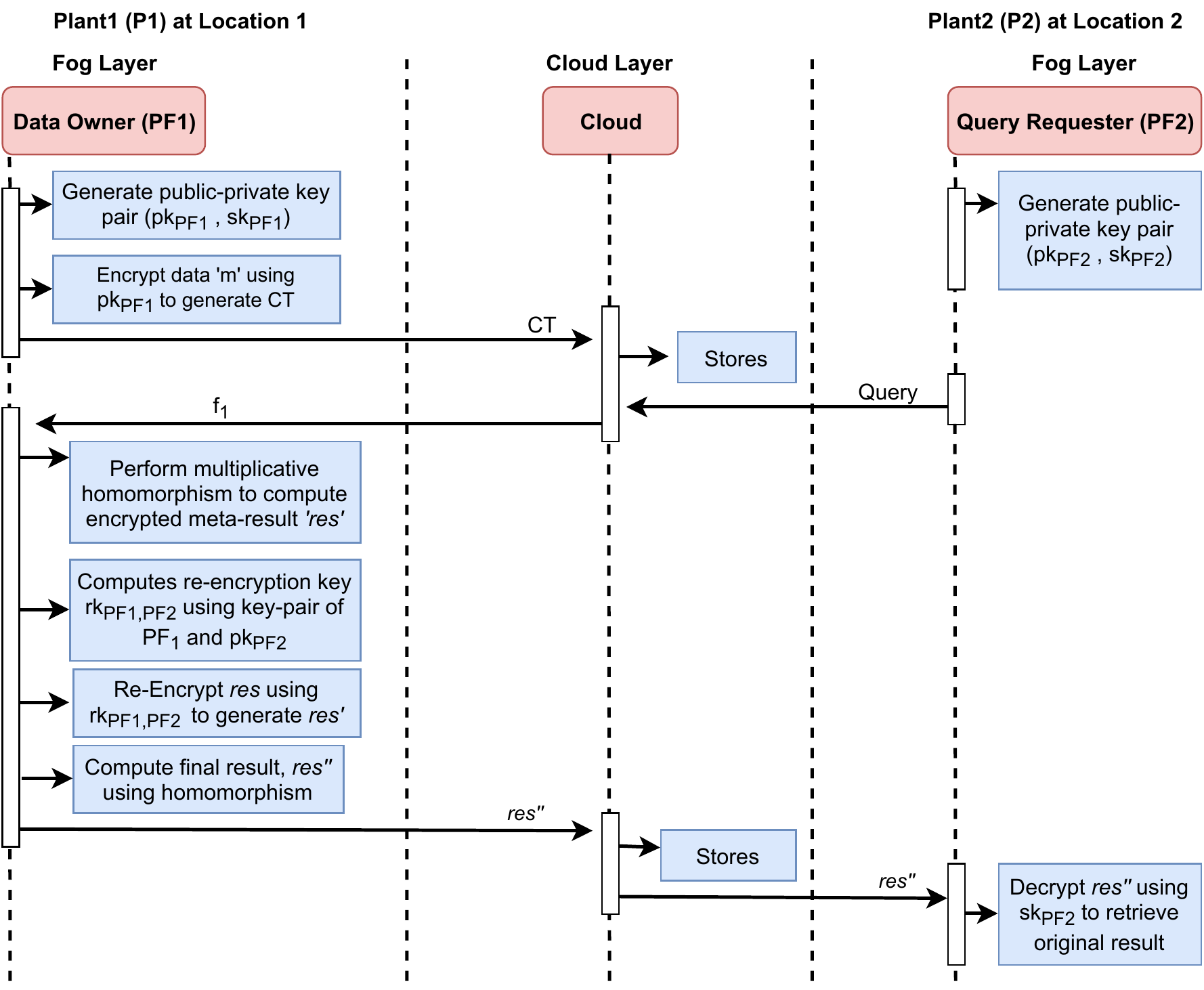}
\caption{\small \sl Sequence Diagram for Secure Computation \label{fig:Image5_1}}
\end{center}  
\end{figure}

The existing techniques of delegating secure computations in literature \cite{c5,e7,e8} propose using a proxy server to perform the tasks on behalf of the cloud. However, in our secure Fog-based IIoT architecture we propose offloading these tasks to the fog nodes which is achieved by modifying one of the existing schemes \cite{c5}. Figure \ref{fig:Image5_1} demonstrates in detail the exact steps performed by the entities in our architecture to achieve secure computation and the interactions between such entities. Let us assume that an industry has two plants ($P_1$ and $P_2$) at different geographic locations. A Fog node acting as a proxy ($PF_1$) in $P_1$ is the data owner who uploads the encrypted data to the Cloud while another Fog node acting as the proxy ($PF_2$) in $P_2$ is the data requester. \smallskip

\noindent \textbf{Setup:} In this phase a Trusted Authority (TA) generates security parameters for the entire industry. It also generates a bilinear map as defined under Section IV.

\noindent \textbf{Key Generation:} Each of the Fog nodes acting as proxies generate their own public-private key pair as per \cite{c5}. Both $PF_1$ and $PF_2$ derive their key pair as $(pk_{PF1}, sk_{PF1})$ = $((pk_{1(PF1)},pk_{2(PF1)}),sk_{PF1})$ and $(pk_{PF2}, sk_{PF2})$ = $((pk_{1(PF2)},pk_{2(PF2)}),sk_{PF2})$ respectively.

\noindent \textbf{Upload Encrypted Data:} Encrypting a message ($m$) under public key $pk_i$ = $(pk_{1i},pk_{2i})$ to generate cyphertext $c=(c_1,c_2)$ takes place as follows: [where $y$ is chosen randomly from $\mathbb{Z}_q$]
$$ c_1=g^y \ ; \ c_2=pk^y_{1i}.m \eqno{...(a)} $$

\noindent Whenever $PF_1$ has a data ($d$) to upload to cloud for statistical or historical analysis purposes, it first encrypts the data using Eq. (a) under its own public key ($pk_{PF1}$) to generate ciphertext $CT=(CT_1,CT_2)$. \smallskip

\noindent When $PF_2$ in $P_2$ generates a query on this encrypted data (CT) to the cloud, the cloud forwards $<CT,f_1>$ to $PF_1$ (where $f_1$ is the evaluation function). $PF_1$ then performs the following four operations before forwarding the final analyzed result back to the cloud. 

\noindent \textbf{Computation on Encrypted Data:} Homomorphic Encryption is used to perform computations on encrypted data (here it only supports multiplicative homomorphism). Given two ciphertexts $ct=(ct_1,ct_2)$ and $ct'=(ct'_1,ct'_2)$, the resulting evaluated ciphertext $c'=(c'_1,c'_2)$ under public key $pk_i$ = $(pk_{1i},pk_{2i})$ is calculated as below: [where $a$ is chosen randomly from $\mathbb{Z}_q$ and $\hat{b}=b+b'+a \ mod \ q$]
$$c'_1=ct_1.ct'_1.g^a=g^{b+b'+a}=g^{\hat{b}}\eqno{...(b)} $$
$$c'_2=ct_2.ct'_2.pk^a_{1i}=pk^{b+b'+a}_{1i}.m.m'=pk^{\hat{b}}_{1i}.m.m'\eqno{...(c)} $$

\noindent $PF_2$ calculates the encrypted meta-result $res=(res_1,res_2)$ for $CT$ and $f_1$ using his/her own public key $pk_{PF1}$ following Equations (b) and (c) above.

\noindent \textbf{Re-Encryption Key Generation:} The re-encryption key $rk_{{PF1},{PF2}}$ using key pair of $PF_1$ and public key of $PF_2$ is calculated as per \cite{c5}.

\noindent \textbf{Re-Encryption:} This phase transforms the meta-result ($res$) encrypted under public key of $PF_1$ to the one encrypted under public key of $PF_2$. Using $res=(res_1,res_2)$ and $rk_{{PF1},{PF2}}$, the transformed result $res'=(res'_1,res'_2)$ is calculated as per \cite{c5}.

\noindent \textbf{Computation on Transformed Result:} Here, $PF_1$ again uses homomorphic encryption to perform analysis on the transformed result using public key of $PF_2$ following Equations (b) and (c). The final result $res''=(res''_1,res''_2)$ is sent to the cloud. The cloud then replies $PF_2$ with $res''$.

\noindent \textbf{Decryption:} $PF_2$ performs decryption of $res''$ under its own private key ($sk_{PF2}$) according to \cite{c5} to retrieve the original result.

\noindent By using this mechanism, the trust on the cloud is completely eliminated. Moreover, computations are performed by Fog nodes which reduces the burden on the constrained industrial machines. Further, the raw data is not exposed and only the encrypted final result is sent to the query requester which greatly secures the system apart from saving network bandwidth.

\section{Performance Evaluation}

In this section we evaluate the performance of our proposed architecture both theoretically and experimentally.

\subsection{Theoretical Analysis}
The computation and communication overheads are measured in terms of execution time and number of transmitting bytes respectively. During analysis, we consider Type A pairings where each group element is of size 128 bytes \cite{i9}. We also consider SHA-256 as the cryptographic hash function with message digest of size 32 bytes. Interested readers can look into \textit{Appendix C} for detailed calculation of overheads. Table \ref{tab:table4} summarizes the notations used.

\begin{table}[!ht]
\centering
\caption{\small \sl Notations used for Theoretical Analysis}
\label{tab:table4}
\scalebox{0.75}{
\begin{tabular}{c|c}
\hline
\rowcolor[HTML]{C0C0C0}
\textbf{Notations} & \textbf{Meaning} \\ \hline
$T_P$ & Time taken to perform a pairing operation \\ \hline
$T_M$ & Time taken to perform a multiplication \\ \hline
$T_H$ & Time taken to perform a hash operation \\ \hline
$T_E$ & Time taken to perform an exponentiation \\ \hline
$T_D$ & Time taken to perform a division \\ \hline
$T_S$ & Time taken to perform a subtraction \\ \hline
$n$ & Number of data packets \\
\hline
$|m|$ & Size of one generated data \\
\hline
$|Req\_{msg}|$ & Size of one request \\
\hline
$x$ & Number of attributes \\
\hline
$l$ & Number of rows in access matrix  $\mathbb{A}$\\
\hline
\end{tabular}
}
\end{table}

\begin{table}[!ht]
\centering
\caption{\small \sl Comparison of Theoretical Overhead Analysis for Secure Data Aggregation}
\label{tab:table5}
\scalebox{0.70}{
\begin{tabular}{|c|c|c|c|c|c|}
\hline
\rowcolor[HTML]{C0C0C0} 
\cellcolor[HTML]{C0C0C0} & \cellcolor[HTML]{C0C0C0} & \multicolumn{2}{c|}{\cellcolor[HTML]{C0C0C0}\textbf{Computation Overhead}} & \multicolumn{2}{c|}{\cellcolor[HTML]{C0C0C0}\textbf{Communication Overhead (bytes)}} \\ \cline{3-6} 
\rowcolor[HTML]{C0C0C0} 
\multirow{-2}{*}{\cellcolor[HTML]{C0C0C0}\textbf{Devices}} & \multirow{-2}{*}{\cellcolor[HTML]{C0C0C0}\textbf{Tasks}} & \textbf{Aggregate Bls} & \textbf{BLS} & \textbf{Aggregate BLS} & \textbf{BLS} \\ \hline
\cellcolor[HTML]{EFEFEF} & Sign & $n(T_E+T_H)$ & $n(T_E+T_H)$ &  &  \\ \cline{2-4}
\multirow{-2}{*}{\cellcolor[HTML]{EFEFEF}\begin{tabular}[c]{@{}c@{}}IoT\\ Device\end{tabular}} & Aggregate & $(n-1)T_H$ & - & \multirow{-2}{*}{$n|m|$ + 96} & \multirow{-2}{*}{$n|m|$ + $n$*96} \\ \hline
\cellcolor[HTML]{EFEFEF}Fog Node & Verify & $nT_H$+$(n+1)T_P$ & $nT_H$+2$nT_P$ & - & - \\ \hline
\end{tabular}
}
\end{table}

\begin{table}[!ht]
\centering
\caption{\small \sl Theoretical Overhead Analysis of Secure Data Sharing with Multiple Users}
\label{tab:table6}
\scalebox{0.75}{
\begin{tabular}{|c|c|c|c|}
\hline
\rowcolor[HTML]{C0C0C0} 
\textbf{Entities} & \textbf{Tasks} & \textbf{\begin{tabular}[c]{@{}c@{}}Computation\\  Overhead\end{tabular}} & \textbf{\begin{tabular}[c]{@{}c@{}}Communication\\  Overhead (bytes)\end{tabular}} \\ \hline
\cellcolor[HTML]{EFEFEF} & Key Generation & 3$T_E$ &  \\ \cline{2-3}
\cellcolor[HTML]{EFEFEF} & Encryption & \begin{tabular}[c]{@{}c@{}}4$T_E$+$T_M$\\ +$T_P$\end{tabular} &  \\ \cline{2-3}
\multirow{-3}{*}{\cellcolor[HTML]{EFEFEF}\begin{tabular}[c]{@{}c@{}}IoT Device\\ (Sender)\end{tabular}} & \begin{tabular}[c]{@{}c@{}}Re-Encryption\\ Key Generation\end{tabular} & \begin{tabular}[c]{@{}c@{}}4$T_E$+$T_H$+\\ $T_M$+$T_P$\end{tabular} & \multirow{-3}{*}{640} \\ \hline
\cellcolor[HTML]{EFEFEF}Fog Node & Re-Encryption & $T_M$+$T_P$ & 384 \\ \hline
\end{tabular}
}
\end{table}

\begin{table}[!ht]
\centering
\caption{\small \sl Theoretical Overhead Analysis for Fine-Grained Access Control with Improved Performance}
\label{tab:table7}
\scalebox{0.75}{
\begin{tabular}{|c|c|c|c|}
\hline
\rowcolor[HTML]{C0C0C0} 
\textbf{Devices} & \textbf{Tasks} & \textbf{\begin{tabular}[c]{@{}c@{}}Computation \\ Overhead\end{tabular}} & \textbf{\begin{tabular}[c]{@{}c@{}}Communication \\ Overhead (bytes)\end{tabular}} \\ \hline
\cellcolor[HTML]{EFEFEF} & \begin{tabular}[c]{@{}c@{}}Intermediate \\ Encryption (At P1)\end{tabular} & \begin{tabular}[c]{@{}c@{}}$x$(9$T_E$+\\ 4$T_M$ +3$T_P$)\end{tabular} & $|m| + x * 640$ (At P1) \\ \cline{2-4} 
\cellcolor[HTML]{EFEFEF} & Key Transform (At P2) & 2$T_E$+$T_H$ &  \\ \cline{2-3}
\multirow{-3}{*}{\cellcolor[HTML]{EFEFEF}IoT Device} & Full Decrypt (At P2) & \begin{tabular}[c]{@{}c@{}}2$T_E$+$T_M$\\ +$T_D$+$T_P$\end{tabular} & \multirow{-2}{*}{160 (At P2)} \\ \hline
\cellcolor[HTML]{EFEFEF} & Full Encrypt (At P1) & \begin{tabular}[c]{@{}c@{}}$T_E$+$T_M$+$T_P$\\ +$x$(4$T_M$+2$T_S$)\end{tabular} & $x$ * 640 \\ \cline{2-4} 
\multirow{-2}{*}{\cellcolor[HTML]{EFEFEF}Fog Node} & Partial Decrypt (At P2) & \begin{tabular}[c]{@{}c@{}}$l$(4$T_E$+3$T_P$+\\ 2$T_D$+$T_M$+$T_H$)\end{tabular} & 256 \\ \hline
\end{tabular}
}
\end{table}

\begin{table}[!ht]
\centering
\caption{\small \sl Theoretical Overhead Analysis for Secure Computation}
\label{tab:table8}
\scalebox{0.75}{
\begin{tabular}{|c|c|c|c|}
\hline
\rowcolor[HTML]{C0C0C0} 
\textbf{Entities} & \textbf{Tasks} & \textbf{Computation Overhead} & \textbf{\begin{tabular}[c]{@{}c@{}}Communication \\ Overhead (bytes)\end{tabular}} \\ \hline
\cellcolor[HTML]{EFEFEF} & Key Generation & 2$T_E$+$T_P$ &  \\ \cline{2-3}
\cellcolor[HTML]{EFEFEF} & Encryption & 2$T_E$+$T_M$ &  \\ \cline{2-3}
\cellcolor[HTML]{EFEFEF} & \begin{tabular}[c]{@{}c@{}}Computation on\\ Encrypted Data\end{tabular} & 4$T_M$+2$T_E$ &  \\ \cline{2-3}
\cellcolor[HTML]{EFEFEF} & \begin{tabular}[c]{@{}c@{}}Re-Encryption\\ Key Generation\end{tabular} & $T_E$ &  \\ \cline{2-3}
\cellcolor[HTML]{EFEFEF} & Re-Encryption & \begin{tabular}[c]{@{}c@{}}2($T_M$+$T_E$\\ +$T_D$)+$T_P$\end{tabular} &  \\ \cline{2-3}
\multirow{-6}{*}{\cellcolor[HTML]{EFEFEF}\begin{tabular}[c]{@{}c@{}}Fog Node\\ (At P1)\end{tabular}} & \begin{tabular}[c]{@{}c@{}}Computation on\\ Transformed Result\end{tabular} & 4$T_M$+2$T_E$ & \multirow{-6}{*}{512} \\ \hline
\cellcolor[HTML]{EFEFEF} & Key Generation & 2$T_E$+$T_P$ &  \\ \cline{2-3}
\multirow{-2}{*}{\cellcolor[HTML]{EFEFEF}\begin{tabular}[c]{@{}c@{}}Fog Node\\ (At P2)\end{tabular}} & Decryption & $T_P$+$T_E$+$T_D$ & \multirow{-2}{*}{$|Req\_{msg}|$} \\ \hline
\end{tabular}
}
\end{table}

\noindent \textbf{Secure Data Aggregation:} The results (Table \ref{tab:table5}) show that the Communication Overhead of an IoT device is significantly less for Aggregate BLS compared to BLS which is achieved at the cost of a little extra computation.

\noindent \textbf{Secure Data Sharing with Multiple Users:} We conclude from the results (Table \ref{tab:table6}) that the task re-encryption ($T_M+ T_P$ time) which is performed by Fog instead of a typical proxy server makes our architecture more trustworthy.

\noindent \textbf{Fine-Grained Access Control with Improved Performance:} The results (Table \ref{tab:table7}) justify that offloading of various tasks to Fog nodes has successfully reduced the computational burden on IoT devices and also eliminated trust on the cloud.

\noindent \textbf{Secure Computation:} We observe from the results (Table \ref{tab:table8}) that the fog nodes perform tasks which would otherwise have been shared among some third party entities. 

\subsection{Experimental Analysis}

\subsubsection{Simulation} We evaluate the performance of Secure Data Aggregation by simulating both BLS and Aggregate BLS. 

\noindent \textbf{Simulation Environment:} We use two computers with different specifications as shown in Table \ref{tab:table1}. Here, one computer with relatively low specification acts as the Industrial Equipment (E) and the other with higher specification acts as the Fog Node (F). We develop two separate Python codes\footnote{\label{note1}\url{https://tinyurl.com/SecureArchitectureCodes}} using the \textit{blspy library} to run in each of the computers. 

\begin{table}[!ht]
\centering
\caption{\small \sl Specifications of the Simulation Environment}
\label{tab:table1}
\scalebox{0.75}{
\begin{tabular}{|c|c|c|}
\hline
\rowcolor[HTML]{C0C0C0} 
\textbf{Specifications} & \textbf{Industrial Equipment (E)}                                                  & \textbf{Fog Node (F)}                                                             \\ \hline
Memory                  & 3.5GiB                                                                             & 7.7GiB                                                                            \\ \hline
Processor               & \begin{tabular}[c]{@{}c@{}}Intel® Core™ i5-7200U \\ CPU @ 2.50GHz * 4\end{tabular} & \begin{tabular}[c]{@{}c@{}}Intel® Core™ i7-6700 \\ CPU @ 3.40GHz * 8\end{tabular} \\ \hline
OS                      & \begin{tabular}[c]{@{}c@{}}64-bit Ubuntu 18.04.3 \\ LTS\end{tabular}               & \begin{tabular}[c]{@{}c@{}}64-bit Ubuntu 18.04.3 \\ LTS\end{tabular}              \\ \hline
Disk                    & 100.3 GB                                                                           & 455.1 GB                                                                          \\ \hline
\end{tabular}
}
\end{table}

\begin{figure} [!ht]
\begin{center}  
\includegraphics[scale=0.30]{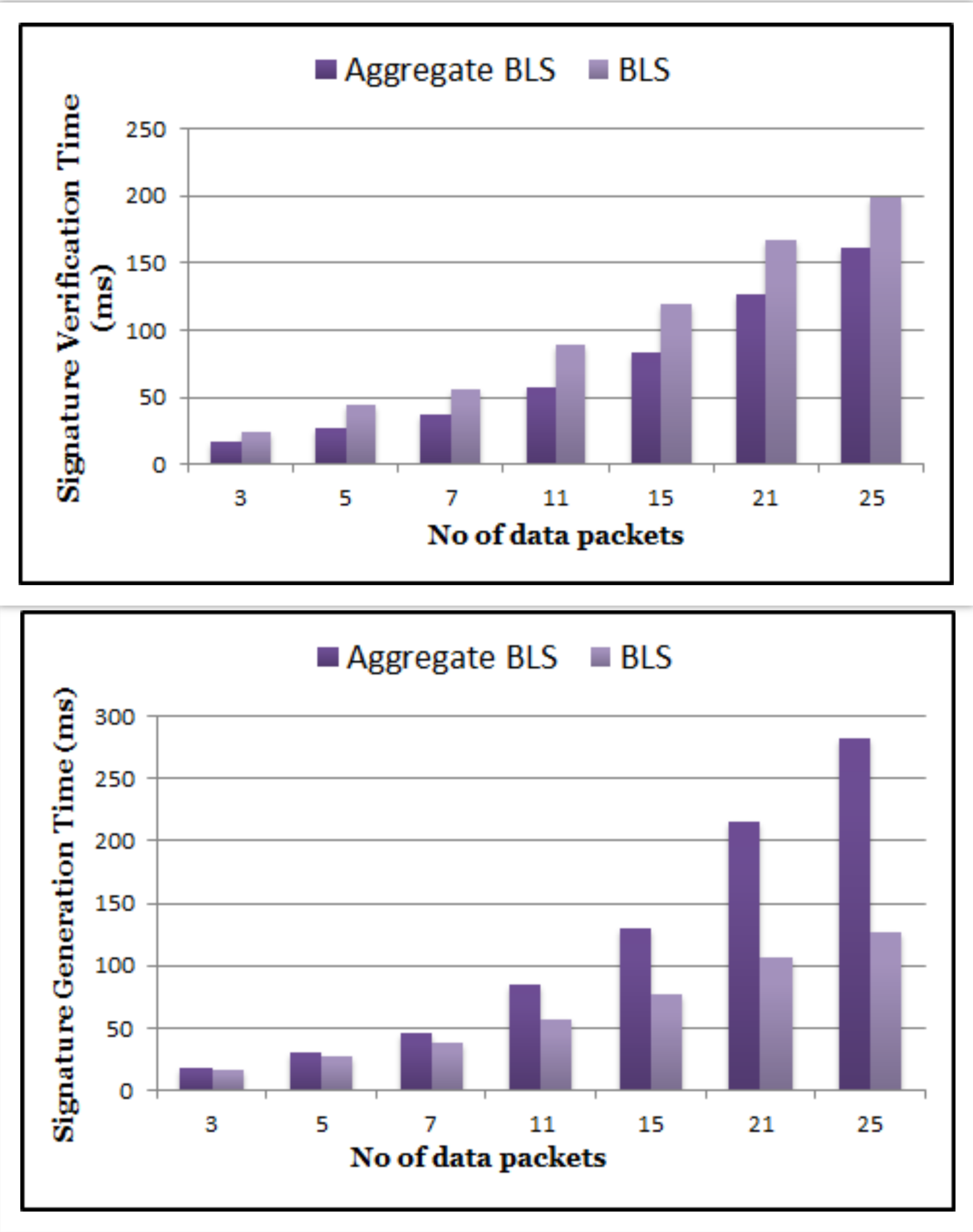}
\caption{\small \sl Comparison of Simulation Results \label{fig:Image6}}
\end{center}  
\end{figure}

\noindent \textbf{Simulation Metrics:} Here, signature generation and verification time are used as metrics. Signature Generation Time is the time taken to generate signatures for individual data packets. Signature Verification Time is the amount of time the fog node takes to verify the signatures of the received data packets. 

\noindent \textbf{Results and discussion:} We conduct two sets of experiments. For each set, the average results of ten independent runs for each input size has been registered.

\noindent In the first set of experiments, we plot (Figure \ref{fig:Image6}) the Signature Generation Time with varying number of data packets. We observe that signature generation time for Aggregate BLS is more than normal BLS. For example, when the number of data packets transferred is 7, Aggregate BLS takes 8.9ms more compared to BLS to generate signatures. In the second set of experiment, we plot (Figure \ref{fig:Image6}) the Signature Verification Time with varying number of data packets. We observe that verification time of Aggregate BLS is less than BLS. For example, when the number of data packets transferred is 7, Aggregate BLS takes 18.9ms less compared to BLS to verify signatures.

\noindent The simulation results clearly conform to the results obtained through theoretical analysis in Table \ref{tab:table5}. Further, the overall execution time of Aggregate BLS is less than BLS.

\subsubsection{Testbed Implementation}

We implement Secure Data Sharing and validate its performance using a testbed setup.

\noindent \textbf{Testbed Setup:} Experimentation is conducted through testbed whose detailed design and description is shown in \textit{Appendix D}. The specifications of devices used in the testbed are shown in Table \ref{tab:table2}. We have developed separate C codes\cref{note1} for each of the devices involved in the setup with the help of \textit{PBC and GMP libraries} to implement our proposed security feature. Through our experimentation we examine the execution times of the tasks performed by each of the devices individually.

\begin{table}[!ht]
\centering
\caption{\small \sl Testbed Setup Specifications}
\label{tab:table2}
\scalebox{0.75}{
\begin{tabular}{|c|c|c|c|}
\hline
\rowcolor[HTML]{C0C0C0} 
\textbf{Specifications} & \textbf{\begin{tabular}[c]{@{}c@{}}Sender \\ (Raspberry Pi)\end{tabular}} & \textbf{\begin{tabular}[c]{@{}c@{}}Fog Node \\ (Laptop)\end{tabular}} & \textbf{\begin{tabular}[c]{@{}c@{}}Data Analyst \\ (Laptop)\end{tabular}} \\ \hline
Memory & 1 GBB & 3.5 GiB & 7.7 GiB \\ \hline
Processor & \begin{tabular}[c]{@{}c@{}}Cortex-A53, armv7l \\ @1200MHz * 4\end{tabular} & \begin{tabular}[c]{@{}c@{}}Intel® Core™ i5-7200U\\ CPU @ 2.50GHz * 4\end{tabular} & \begin{tabular}[c]{@{}c@{}}Intel® Core™ i7-6700\\ CPU @ 3.40GHz * 8\end{tabular} \\ \hline
OS & 32-bit Raspbian & \begin{tabular}[c]{@{}c@{}}64-bit Ubuntu \\ 18.04.3\end{tabular} & \begin{tabular}[c]{@{}c@{}}64-bit Ubuntu\\ 18.04.3\end{tabular} \\ \hline
Disk & 16 GB & 100.3 GB & 455.1 GB \\ \hline
\end{tabular}
}
\end{table}

\begin{table}[!ht]
\centering
\caption{\small \sl Experimental Results}

\label{tab:table3}
\scalebox{0.80}{
\begin{tabular}{|c|c|}
\hline
\rowcolor[HTML]{C0C0C0} 
\textbf{Tasks} & \textbf{Execution Time (in ms)} \\ \hline
PKG Setup & 9.964 \\ \hline
Key Generation  (Sender) & 30.317 \\ \hline
Key Generation (Receiver) & 2.311 \\ \hline
Encryption & 57.717 \\ \hline
Re-Encryption Key Generation & 45.997 \\ \hline
Re-Encryption & 0.723 \\ \hline
Decryption & 0.581 \\ \hline
\end{tabular}
}
\end{table}

\noindent \textbf{Results and discussion:} In our experiment, Type A pairings are used which are constructed on the curve $y^2=x^3+x$ over the field $\mathbb{F}_q$  for some prime q = 3 mod 4 with base field size of 512 bits. The average execution times obtained after ten independent runs for each task are shown in Table \ref{tab:table3}.

\noindent We observe from the results obtained in Table \ref{tab:table3} that the execution times of each of the tasks lie within a feasible range and is suitable for IIoT scenarios. This conforms to the results obtained through theoretical analysis in Table \ref{tab:table6}. Moreover, re-encryption which has been delegated to the fog nodes also takes almost negligible time but provides more trustworthiness.

\subsection{Metrics of Success}

We highlight the metrics of success based on the performance evaluation results of our proposed architecture. The detailed description is available in \textit{Appendix E}.

\begin{itemize}[leftmargin=*]
    \item \textbf{More Secure:} Plugging and playing a number of security feature in an integrated manner while keeping overheads within an acceptable limit.
    \item \textbf{Reduced Overhead on Low End Devices:} Applying appropriate securing scheme and offloading tasks from low end devices to fog nodes suitably.
    \item \textbf{Reduced Latency:} Offloading major tasks judiciously to the powerful fog nodes in general and applying appropriate securing schemes in particular.
    
\end{itemize}

\section{Conclusion}

With the rising popularity of IIoT and Industry 4.0, several limitations both in terms of networking and security/privacy have emerged. Thus, we incorporate a few modified security features to fit into our proposed Fog based IIoT architecture. Through each of these features, we have shown how the trust dependency on the cloud gets eliminated while successfully using it for storage and archival purpose. Further, we have shown that by offloading several computationally intensive tasks to the fog nodes, the battery life of the resource-constrained end devices is greatly saved. This architecture also minimizes the latency in decision making thereby improving performance. Finally, we have validated the performance of our proposed architecture through theoretical analysis and practical implementation. In future, we are planning to incorporate additional security features to develop a more realistic architecture suitable for industrial scenarios. 

\appendices

\section{Exemplary Industrial Scenarios}

Here, we describe a couple of exemplary industrial scenarios to justify the need of including various security features into our proposed Fog based IIoT architecture.

\begin{itemize}
    \item An industry may decide not to allow the workers in the inventory to access data in the production floor and vice-versa. To incorporate this, the industries require proper access control mechanisms. 
    \item The data packet containing sensed temperature data (say above normal temperature) from a specific machine gets intercepted by an adversary and is tampered (to normal temperature) before being delivered to the nearest fog node. In such a case, the fog node being completely unaware of the situation will not take the necessary actions leading to damage of the machine. To handle such unprecedented situations, protecting the confidentiality and integrity of raw data in transit is a necessity.
\end{itemize}

\section{Bilinear Maps and Pairing Concept}

A bilinear map is a map $\hat{e} : \mathbb{G}_1 \times \mathbb{G}_1 \rightarrow \mathbb{G}_T$ where $\mathbb{G}_1$ is a Gap Diffie-Hellman (GDH) group. The group $\mathbb{G}_1$ is a subgroup of the additive group of points of an elliptic curve $E/\mathbb{F}_p$. The group $\mathbb{G}_T$ is a
subgroup of the multiplicative group of a finite field $\mathbb{F}^*_{p^2}$ \cite{i7}. The map has the following properties \cite{i7}:

\begin{itemize}
\item \textbf{Bilinear:} For all P, Q $\in$ $\mathbb{G}_1$ and for $\forall$ c,d $\in$ $\mathbb{Z}^*_q$, we have:
$$\hat{e}(cP,dQ) = \hat{e}(cP,Q)^d = \hat{e}(P,dQ)^c = \hat{e}(P,Q)^{cd}$$

\item \textbf{Non-degenerate:} If P is a generator of $\mathbb{G}_1$, then the following happens:
$$\forall P \in \mathbb{G}_1, P \neq 0 \Rightarrow \ \hat{e}(P,P) = \mathbb{G}_2 \ (\hat{e}\ (P,P)$$ generates  $\mathbb{G}_2)$

\item \textbf{Computable:} There is an efficient algorithm to compute $\hat{e}(P,Q)$  $\forall \ P,Q \in \mathbb{G}_1$. 
\end{itemize}

\section{Theoretical Overhead Analysis}

In this part, we discuss in detail the computation and communication overheads for each of the schemes mentioned in Section IV.

\subsection{Secure Data Aggregation}

\noindent \textbf{Computation Overhead:} In this scheme, the computation overhead of an IoT device for signing a single data packet is ($T_E+T_H$) and is same for both Aggregate BLS and BLS. For $n$ number of data packets it will be $n(T_E+T_H)$. In Aggregate BLS, an additional time is required to aggregate $n$ signatures and is equal to $(n-1)T_H$. 

\noindent \textbf{Communication Overhead:} Contrary to the above discussion, the communication overhead of an IoT device is significantly less in case of Aggregate BLS compared to BLS. Here, IoT device has to send only a single aggregated signature of 96 bytes [16] along with the $n$ data packets with size $|m|$. Thus, communication overhead for Aggregate BLS is equal to $n|m| + 96$ bytes. However, in case of BLS, the device has to send $n$ signatures of 96 bytes each in addition to $n$ data packets with size $|m|$.

\noindent We observe from the results that the Communication Overhead of an IoT device is significantly less for Aggregate BLS compared to BLS. We achieve this at the cost of a little extra computation in Aggregate BLS which is on the basis of execution time. We also observe that for a Fog Node, Computation Overhead is less in case of Aggregate BLS compared to BLS. However, for BLS it increases significantly with number of data packets. Thus, Aggregate BLS is advantageous to implement in our Fog based IIoT architecture.

\subsection{Secure Data Sharing with Multiple Users}

\noindent \textbf{Computation Overhead:} To implement the access control scheme, the computation overhead of IoT device is the resultant computation overhead for Key Generation, Encryption and Re-Encryption Key Generation.\\
Key Generation = $T_E$ (private key) + $T_E$ (public key) + $T_E$ ($g^d$) = 3$T_E$\\
Encryption = $T_E$ ($c_0$) + $T_E$ ($c_1$) + ($2T_E+T_M+T_P$) (for $c_2$) = $4T_E+T_M+T_P$\\
Re-Encryption Key Generation = ($2T_E+T_H$) (for $c_4$) + ($2T_E+T_M+T_P$) (for $C_R(y)$) = $4T_E+T_H+T_M+T_P$

\noindent \textbf{Communication Overhead:} As explained under Section IV-C, an IoT device sends a 2-tuple packet $<c',rk_{S,R}>$ to a Fog node. The communication overhead depends on the size of each of these tuples. \\
$c'$ = {128 ($c_0$) + 128 ($c_1$) + 128 ($c_2$) }bytes= 384 bytes\\
$rk_{S,R}$  = 128 bytes ($c_4$) +  128 bytes ($C_R(y)$) = 256 bytes\\
Therefore, total communication overhead = (384 + 256) bytes = 640 bytes

We conclude from the above discussion that the task re-encryption which takes $T_M+ T_P$ time was originally supposed to be performed by the proxy server. By offloading this task to the Fog node we make the architecture more trustworthy.

\subsection{Fine-Grained Access Control with Improved Performance}

\noindent \textbf{Computation Overhead:} To implement this feature, the computation overhead of IoT device at Plant2 (P2) is due to the Key Transformation and Full Decrypt.\\
Key Transformation = ($2T_E+T_H$) (for transformed key)\\
Full Decrypt = $T_E$ ($CT'_2$) + $T_M$ ($CT$) + ($T_E+T_P$) (for $CT^r$) + $T_D$ ($d$) = $2T_E+T_M+T_D+T_P$

\noindent \textbf{Communication Overhead:} The IoT device at Plant2 (P2) sends the transformed key to the Fog Node which is the communication overhead of the IoT device.\\
Transformed Key = 128 bytes (part of the key) + 32 bytes (hashing) = 160 bytes

\noindent In the absence of a Fog layer, Full Decrypt at Plant1 (P1) would have been performed by an IoT device which clearly justifies that our proposed security feature has successfully reduced the computational burden on the resource constrained end devices. Moreover, Partial Decrypt at Plant2 (P2) is performed by a fog node which otherwise would have been performed by a proxy server in the Cloud. Thus, our proposed architecture also eliminates trust on the Cloud by offloading major computations to the fog nodes.

\subsection{Secure Computation}

\noindent \textbf{Computation Overhead:} The computation overhead of Fog Node at Plant2 (P2) for this scheme arises from Key Generation and Decryption.\\
Key Generation = ($2T_E+T_P$) (for Key-pair)\\
Decryption = $T_P+T_E+T_D$

\noindent \textbf{Communication Overhead:} The communication overhead of Fog Node at Plant1 (P1) for this scheme is because of the transmission of $CT$ and $res''$.\\
$CT$ = 128 bytes ($CT_1$) + 128 bytes ($CT_2$) = 256 bytes\\
$res''$ = 128 bytes ($res''_1$) + 128 bytes ($res''_2$) = 256 bytes\\
Total communication overhead = 256 bytes + 256 bytes = 512 bytes

\noindent Thus, we can conclude that in the absence of the fog nodes these tasks would have been performed on sharing information among some third party entities (e.g. gateway, cloud). Thus, we can claim that our proposal mostly eliminates trust on such third party entities except for storage and archival purposes.

\section{Testbed Design}

\begin{figure} [!ht]
\begin{center}  
\includegraphics[scale=0.35]{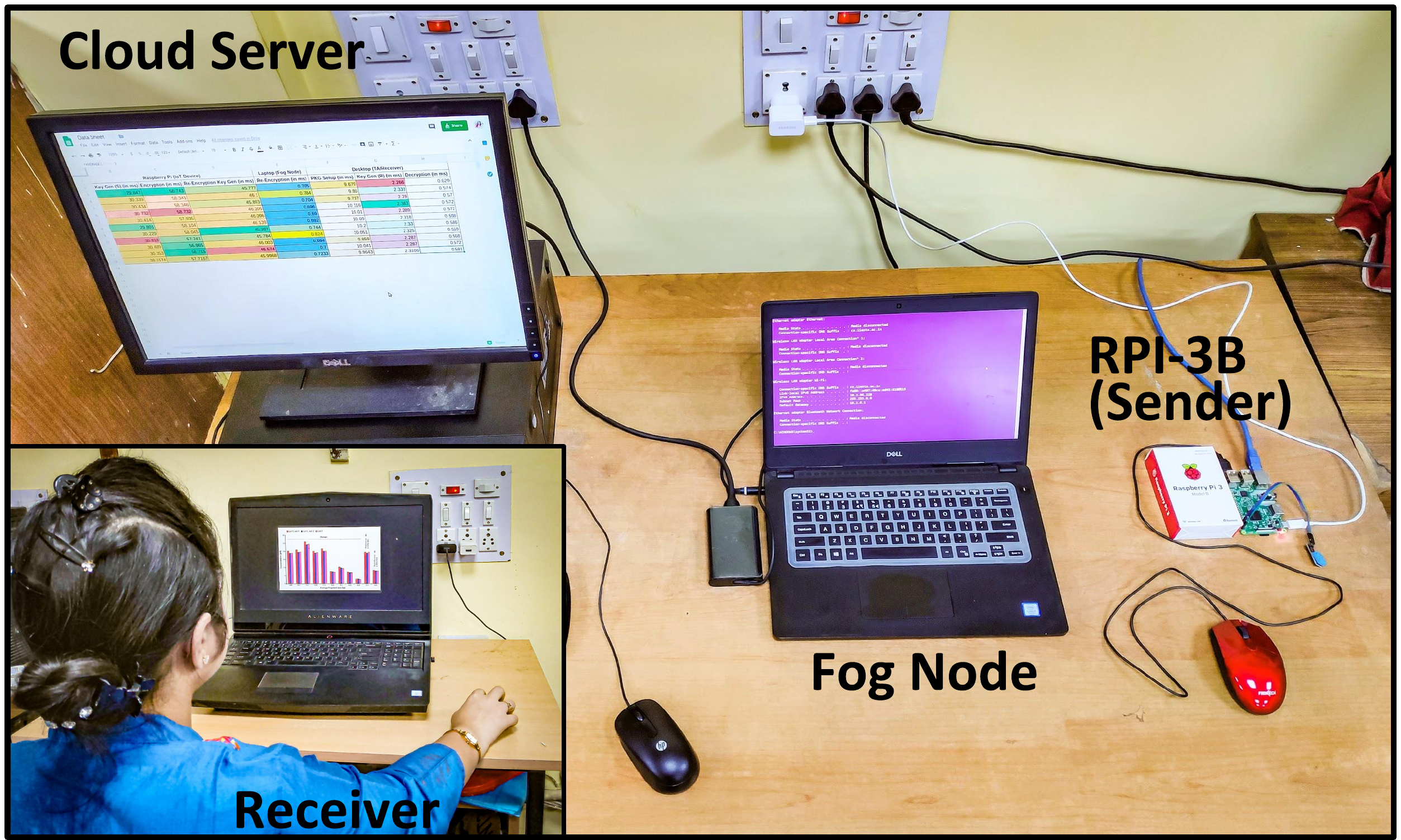}
\caption{\small \sl Testbed Setup of Secure Data Sharing with Multiple Users \label{fig:Image7}}
\end{center}  
\end{figure}

We conduct experimentation through testbed where Raspberry Pi (RPI-3B) is the IoT device acting as the sender.  A Laptop acts as the Fog Node and ThinkSpeak which is a cloud server is running on a desktop. Finally a user is operating at a remote terminal to retrieve data from the system. Figure \ref{fig:Image7} shows our testbed design. 

\section{Metrics of Success}

We highlight the metrics of success based on the performance evaluation results of our proposed secure fog based architecture. 

\begin{itemize}[leftmargin=*]
    \item \textbf{More Secure:} To the best of our knowledge, there hasn't been any such works reported which develops a Secure Fog based architecture for IIoT and Industry 4.0. Plugging and playing each of the individual features separately becomes a tedious process, specifically in industrial environments where huge amounts of data processing takes place at any given time instance and one or more of these features maybe required to be applied. We have been able to successfully secure the architecture at the cost of appreciably acceptable communication, computation overheads and execution time discussed in Tables II-V, VIII respectively (Section V).
    \item \textbf{Reduced Overhead on Low End Devices:} By choosing the appropriate security schemes and applying them intelligently has also helped us to minimise the communication overhead of the resource-constrained end devices. For example, in case of \textit{Secure Data Aggregation}, the communication overhead of Aggregate BLS is comparatively lesser than BLS as a single aggregated signature of size 96 bytes is to be transferred. Thus, we have achieved preferably low communication overhead at the cost of a little larger computation overhead.
    Also by offloading the tasks to the fog nodes we have been able to considerably reduce the computational burden on the resource-constrained devices at the perception layer. For example in the said security feature, \textit{Fine-Grained Access Control with Improved Performance}, in the absence of a Fog layer, Full Decrypt (At P1) would have been performed by an IoT device which clearly justifies that our proposal has successfully reduced the computational burden on the resource constrained end devices. Additionally by reducing the computational burden on the low-end devices, the latency in determining the output also reduces.

    \item \textbf{Reduced Latency:} By applying appropriate schemes and offloading major tasks judiciously to the more powerful Fog nodes, the latency in decision making can be considerably reduced. For example, in case of Secure Data Aggregation, a Fog node may receive data packets for signature verification from multiple industrial equipments at the same time. In such a scenario, the Fog Node has to verify a single signature in case of Aggregate BLS compared to BLS where it has to verify each of the received signatures. This greatly reduces the latency in decision making.
    
\end{itemize}

\bibliographystyle{unsrt}
{\footnotesize
\bibliography{references}}

\begin{IEEEbiographynophoto}{Jayasree Sengupta}
is currently pursuing her PhD in the Department of Computer Science and Technology at Indian Institute of Engineering Science and Technology, Shibpur, India. Previously, she has received her MTech degree in Distributed and Mobile Computing from Jadavpur University, Kolkata, India in 2017. She has published research articles in reputed peer reviewed journals as well as in International conference proceedings. Her research interests include Network Security, Fog computing, Applied Cryptography, Blockchains, IoT and Industrial IoT.
\end{IEEEbiographynophoto}

\begin{IEEEbiographynophoto}{Sushmita Ruj}
received her B.E. degree in Computer Science from Bengal Engineering and Science University, Shibpur, India and Masters and PhD in Computer Science from Indian Statistical
Institute. She was an Erasmus Mundus Post Doctoral Fellow at Lund University, Sweden and Post Doctoral Fellow at University of Ottawa, Canada.
She is currently a Senior Research Scientist at CSIRO Data61, Australia. She is also an Associate Professor at Indian Statistical Institute, Kolkata. Her research interests are in Blockchains, Applied Cryptography, and Data Privacy. She serves as a reviewer of Mathematical Reviews, Associate
Editor of Elsevier Journal , Information Security and Applications and is involved with a number of conferences as Program Co Chairs or committee members. She is a recipient of Samsung GRO award, NetApp Faculty Fellowship, Cisco Academic Grant and IBM OCSP grant. She is a Senior
Member of the ACM and IEEE.
\end{IEEEbiographynophoto}


\begin{IEEEbiographynophoto}{Sipra Das Bit}
is a Professor of the Department of Computer Science and Technology, Indian Institute of Engineering Science and Technology, Shibpur, India. A recipient of the Career Award for Young Teachers from the All India Council of Technical Education (AICTE), she has more than 30 years of teaching and research experience. Professor Das Bit has published many research papers in reputed journals and refereed International conference proceedings. She also has three books to her credit. Her current research interests include Internet of things, wireless sensor network, delay tolerant network, mobile computing and network security. She is a Senior Member of IEEE.
\end{IEEEbiographynophoto}

\end{document}